\documentclass[sigconf]{acmart}
\AtBeginDocument{%
  \providecommand\BibTeX{{%
    \normalfont B\kern-0.5em{\scshape i\kern-0.25em b}\kern-0.8em\TeX}}}

\setcopyright{ccbysa} 
\copyrightyear{2022}
\acmYear{2022}

\acmConference[CHI '22]{CHI Conference on Human
Factors in Computing Systems}{April 29--May 5, 2022}{New Orleans, LA, USA}
\acmBooktitle{CHI Conference on Human Factors in Computing Systems (CHI '22), April 29--May 5, 2022,
New Orleans, LA, USA}

\usepackage{color}
\usepackage{hyperref}
\usepackage{caption}
\usepackage{subcaption}

\begin{document}

\title{Immersive Visual Analysis of Cello Bow Movements}


\settopmatter{authorsperrow=3}


\author{Frank Heyen}
\authornote{All authors contributed equally to this work.}
\email{frank.heyen@visus.uni-stuttgart.de}
\orcid{0000-0002-5090-0133}
\author{Yannik Kohler}
\authornotemark[1]
\email{st156038@stud.uni-stuttgart.de}
\affiliation{%
  \institution{VISUS, University of Stuttgart}
  \city{Stuttgart}
  \country{Germany}
}

\author{Sebastian Triebener}
\authornotemark[1]
\email{sebastian@triebener.de}
\affiliation{%
  \institution{Hochschule für Musik und Darstellende Kunst Stuttgart}
  \city{Stuttgart}
  \country{Germany}}

\author{Sebastian Rigling}
\authornotemark[1]
\email{sebastian.rigling@visus.uni-stuttgart.de}
\orcid{0000-0002-2732-6276}
\author{Michael Sedlmair}
\authornotemark[1]
\email{michael.sedlmair@visus.uni-stuttgart.de}
\orcid{0000-0001-7048-9292}
\affiliation{%
  \institution{VISUS, University of Stuttgart}
  \city{Stuttgart}
  \country{Germany}
  }

\renewcommand{\shortauthors}{Heyen, et al.}

\begin{abstract}
  We propose a 3D immersive visualization environment for analyzing the right hand movements of a cello player.
  To achieve this, we track the position and orientation of the cello bow and record audio.
  As movements mostly occur in a shallow volume and the motion is therefore mostly two-dimensional, we use the third dimension to encode time.
  Our concept further explores various mappings from motion and audio data to spatial and other visual attributes.
  We work in close cooperation with a cellist and plan to evaluate our prototype through a user study with a group of cellists in the near future.
\end{abstract}



\begin{teaserfigure}
   \centering
     \begin{subfigure}[b]{0.325\linewidth}
         \centering
         \includegraphics[width=\linewidth]{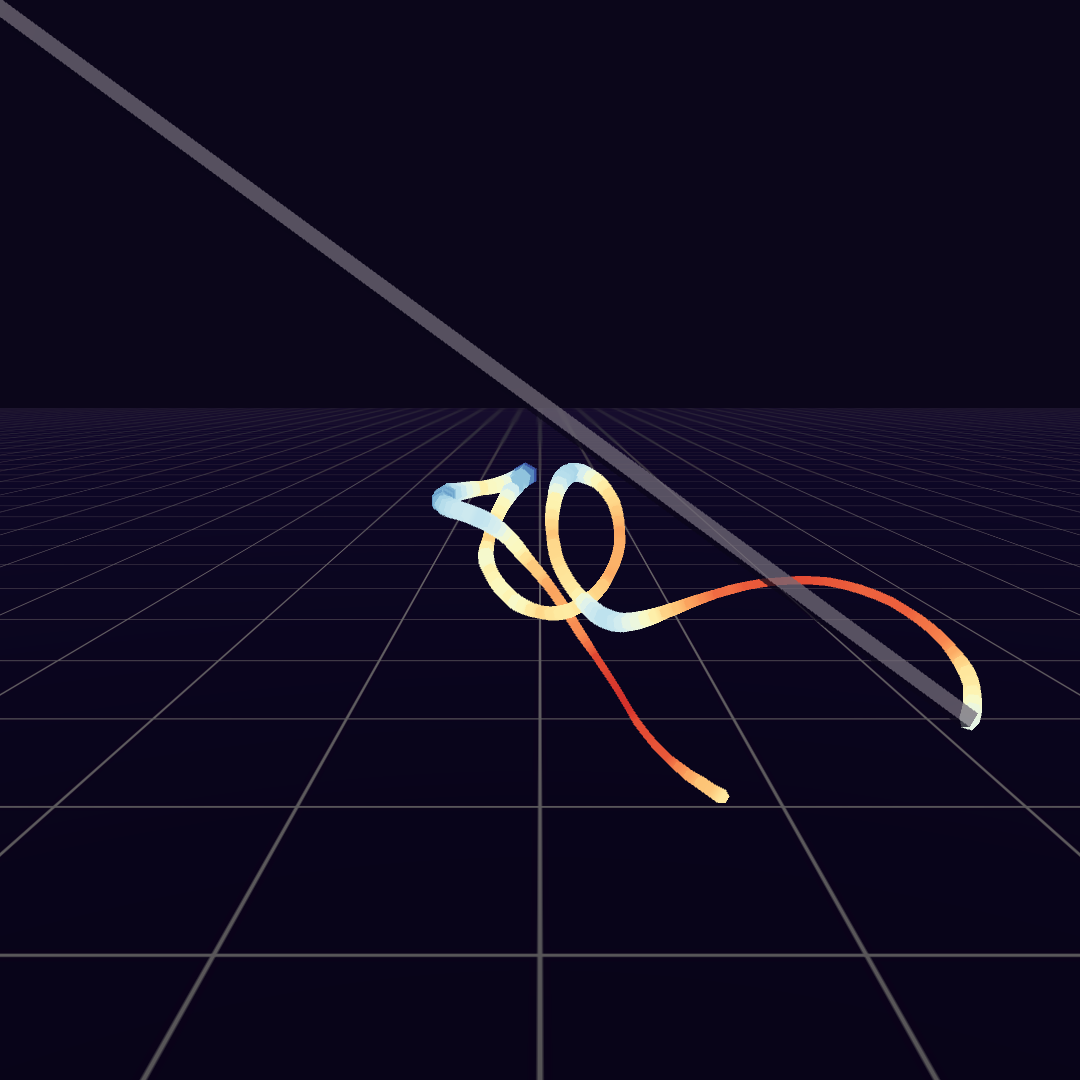}
         \caption{
            A one-to-one mapping that shows the motion exactly as it was.
         }
         \label{fig:one2one}
     \end{subfigure}
     \hfill
     \begin{subfigure}[b]{0.325\linewidth}
         \centering
         \includegraphics[width=\linewidth]{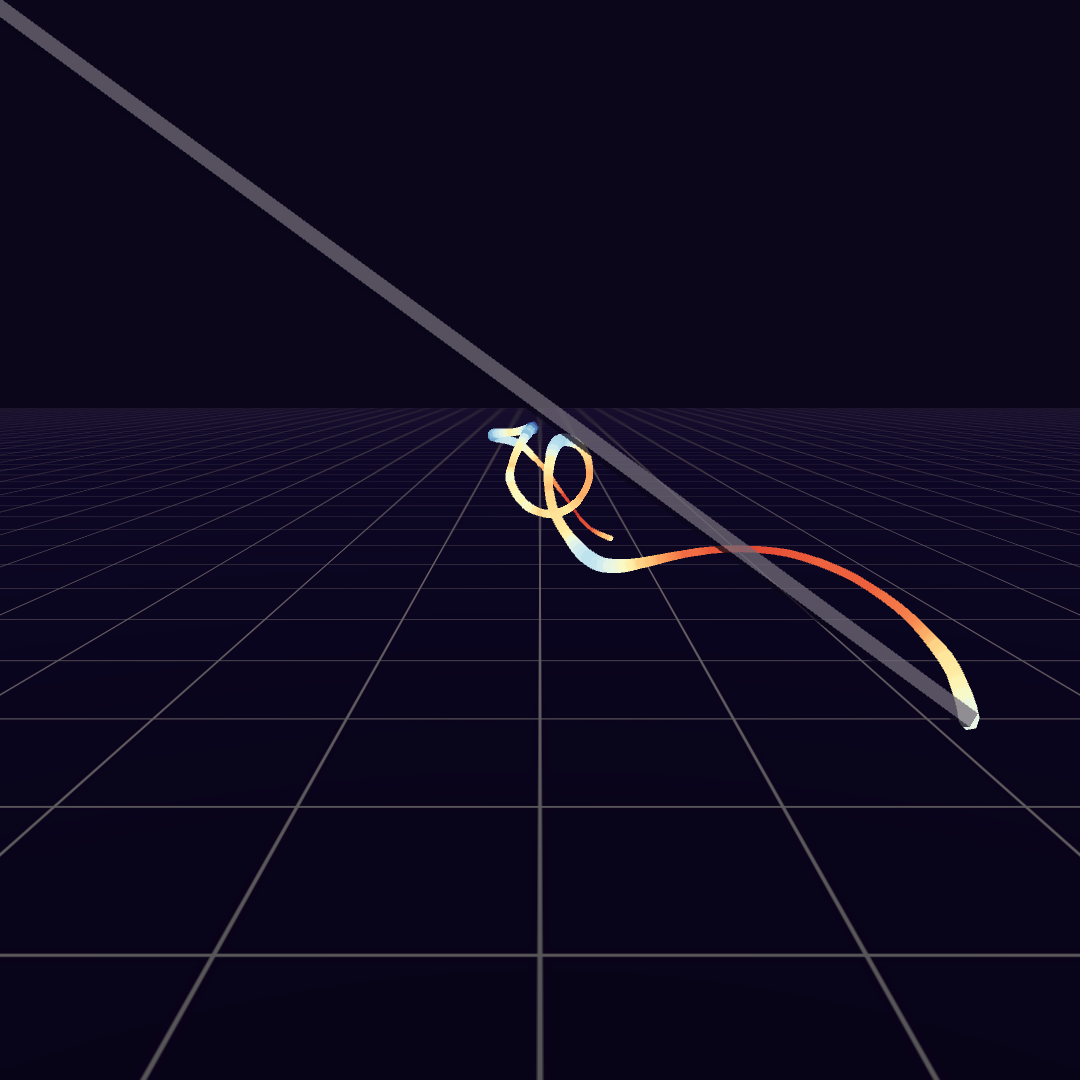}
         \caption{
            Mapping time to a spatial dimension allows seeing longer recordings less cluttered.
         }
         \label{fig:time2space}
     \end{subfigure}
     \hfill
     \begin{subfigure}[b]{0.325\linewidth}
         \centering
         \includegraphics[width=\linewidth]{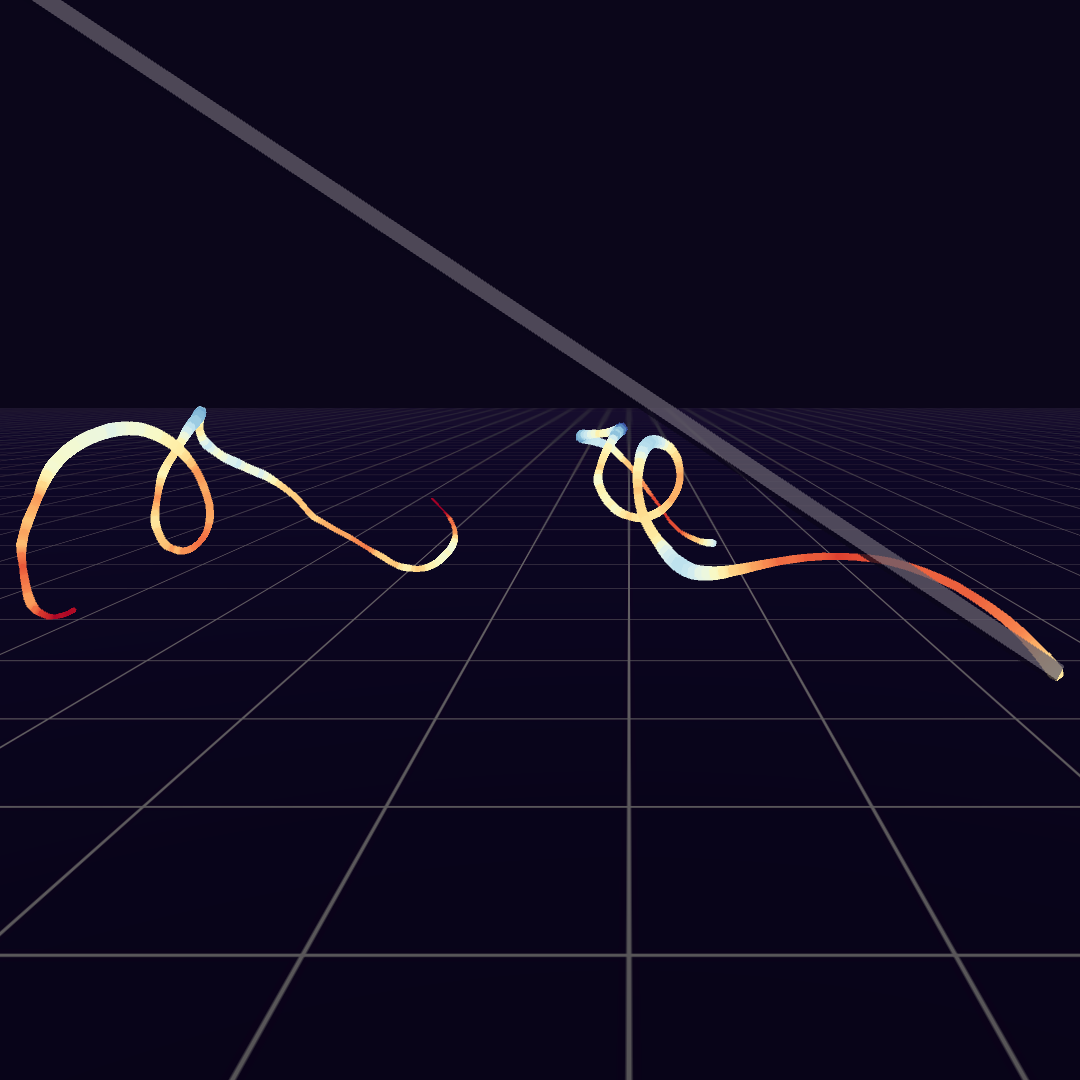}
         \caption{
            Simultaneous display of the current motion and a recording.
         }
         \label{fig:comparison}
     \end{subfigure}
          \caption{
            Comparison of visual encodings for bow movements.
            Here, color and thickness both indicate movement speed. 
            We plan to also visualize played notes, loudness, and differences between recordings or live input and recording.
          }
        \Description{
        Comparison of three examples for visual encodings.
        In all three images, a long gray bar represents the current position and orientation of the cello bow.
        From left to right:
        a)
        A 3D path that shows the last few seconds of hand/bow movement.
        The color and thickness varies and encodes the movement speed, thin and red for fast, thick and blue for slow.
        b)
        The same path as in a), but stretched away from the observer and therefore more spaced out.
        c)
        The same path as in b), with another path next to it, that looks roughly similar.
        }
        \label{fig:teaser}
\end{teaserfigure}

\maketitle

\section{Introduction}

Sheet music for cello pieces usually lacks instructions for how notes should be played.
Deciding on the exact bow movements is therefore up to the musicians themselves.
Since there might be very different ways to play the same piece, the analysis and comparison of different musicians' playing can reveal their strategies and differences and might help musicians try other movements and thereby improve or vary their playing. 

Instrument playing is a physical activity in which positions and scales of movements have a meaning connected to our physical world.
We therefore decided to use concepts from immersive and situated analytics for our design~\cite{ImAnaBook}.
When looking at the bow movements of a cellist, we note that they are mostly restricted to a two-dimensional plane.
This fact can be leveraged for visualization, as it allows us to map time to space, following an important visualization principle~\cite{ware2006whalevis}. 
We thus use the third dimension by letting a line indicating the path of motion ``fly away'' from the observer.

Our approach works for both live and post hoc analysis and can show multiple recordings of the same or different players.
Besides analysis, our display could be used for artistic performances, where the audience sees our 3D representation of the musicians' movements over time.
Our current design uses virtual reality for ease of development, but future work will explore extensions to augmented reality, which would allow musicians and audience to see both instruments and visualization in the same space.

The authors of this paper have different backgrounds in the areas of visualization, immersive analytics, and general human-computer interaction.
During our design, we collaborate closely with a cellist and follow guidelines from design study methodology~\cite{sedlmair2012design}, such as committing to regular meetings with this domain expert.

In summary, we contribute a new interface for immersive analysis of cellists' bow movements to allow studying and comparing movement patterns of one or more musicians.
We explore possible data features we can measure or compute and then map to a range of visual markers and encodings.
Our current prototype supports data collection and basic visualization.
Future work includes further design exploration and an evaluation with a group of cellists.

\section{Related Work}

A recent survey on music visualization~\cite{khulusi2020survey} shows that despite a range of different applications such as visualization of instrument hardware, audio~\cite{foote1999visualizing}, sheet music~\cite{miller2019framing, miller2019augmenting, fuerst2020augmenting}, artist networks, and listening histories~\cite{baur2010streams}, there has still been little work in this domain.
One task that has not been addressed much in this area is the analysis and visual comparison~\cite{gleicher2011visual} of 
motion and exercise recordings.
As data collected from playing music is personal to the musician, we consider our approach to be personal visualization~\cite{huang2015personal:vis:and:va}.

Closest to our work are motion guidance visualizations. 
These systems show animated 3D paths that users need to follow along when learning motions such as dancing~\cite{yu2020perspective} or physiotherapy exercises~\cite{tang2015physio}. 
Our primary tasks are more analytical, and as such we are interested in visualizations that can still provide motion guidance but at the same time allow for a better overview over entire motion sequences.
To that end, we found inspiration in Ware et al.'s work~\cite{ware2006whalevis}, in which they --- instead of using animation --- map the 3D movements and behavior of whales onto paths that are augmented with textures and glyphs.

Work in the field of human-computer interaction addressed tasks such as assessing musicians' playing and providing respective feedback.
Augmented reality, for instance, has been used to display information on drum kits~\cite{yamabe2011feedback} and guitars~\cite{loechtefeld2011guitar}.
\emph{Let's frets!}~\cite{marky2021letsfrets} uses LEDs and touch sensors on the fretboard to display and check exercises.
Other approaches~\cite{karolus2018emguitar,yuksel2016bach} use sensor measurements in order to adapt the tempo or difficulty of exercises, or to allow users to control effects~\cite{karolus2020thumb}.
Colored sheet music~\cite{asahi2018toward:piano:support, hori2019piano:hmm, smith2008interactive} can highlight mistakes and compare student's and teacher's recordings. 
\emph{Strummer}~\cite{ariga2017strummer} teaches users to play guitar chords and uses audio to check for correctness.
\emph{Soloist}~\cite{wang2021soloist} creates overviews of instructional videos for guitar players and extracts the played notes from audio through \emph{CREPE}~\cite{kim2018crepe}, so users can visually compare their playing to the instructor's. 
Our work focuses on different tasks, data, and instruments and extends this set of previous work.

\begin{figure}[tp]
  \centering
  \includegraphics[width=0.65\linewidth]{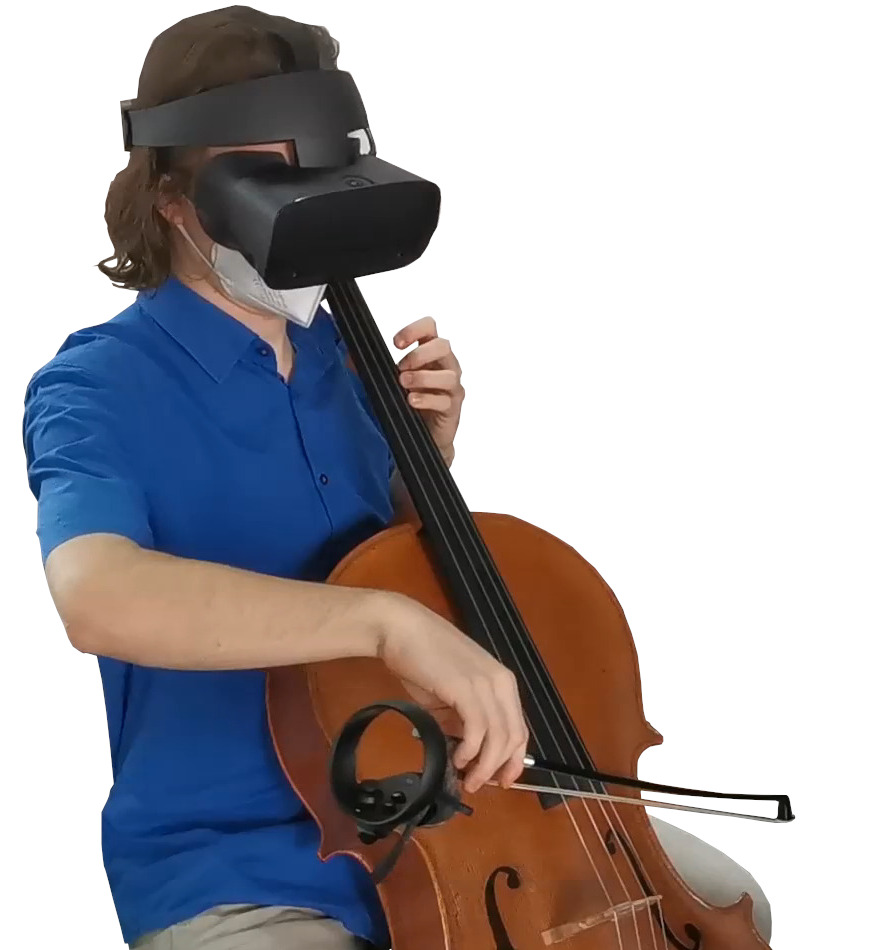}
  \caption{
    We currently track position and orientation of the bow through a VR controller fixed to it. 
  }
  \Description{
  A cellist wears a head-mounted display and plays the cello using a bow, which has a VR controller fixed to it.
  }
  \label{fig:tracking}
\end{figure}

\section{Design \& Implementation}

Inspired by Munzner's nested model~\cite{munzner2009nested}, we structure this section into use cases and tasks, tracking and data, and visualization.
Our current design uses virtual reality (VR) to allow for more immersion~\cite{ImAnaBook} and better embedding~\cite{Willet2017} than desktop interfaces, and a less distracting environment~\cite{satkowski2021investigating} and easier development than an augmented reality (AR) interface.
We display sheet music virtually and our domain expert confirmed that seeing the instrument is not important.
An extension to AR is planned for future work.

\subsection{Use Cases \& Tasks}

There are multiple educational use cases that we plan to support:
In music schools, our approach could be used to record a teacher's motion that a student then plays back while trying to follow, to learn the recommended movements.
Recordings could also be analyzed and compared retrospectively, for instance to find incorrect or very different sections.
Students or self-taught learners who own VR hardware could also do this at home once recordings made by others are available.
Without access to others' recordings, students could still analyze their own playing and detect possible difficulties.

There are also uses cases apart from education:
During a live performance, the musicians' movements could be visualized for the audience, for example through individual head-mounted displays or a 3D cinema canvas and glasses.
Once recorded, the movements themselves could be displayed as an art installation, played back together with audio, and interacted with by visitors.

\subsection{Tracking \& Data}

Our implementation uses the \href{https://unity.com/}{Unity game engine} for rendering 3D real-time visualization and an \href{https://www.oculus.com/}{Oculus Rift S} headset with controllers for output and interaction.
For our current prototype, we taped one of the usually hand-held controllers to the bow (\autoref{fig:tracking}), but we plan to switch to \href{https://www.optitrack.com/}{OptiTrack} for our user study to reduce the bow's weight.
We track the controller's 3D position and orientation 80 times per second (one sample each 12.5 ms) and store this information together with a time stamp.
Additionally, we record audio and extract played notes with CREPE~\cite{kim2018crepe}.

\autoref{tab:data} summarizes the data features that we collect or compute for each time frame.
For note and loudness, we can also compute the difference between actual and expected and visualize this value.
In case we have multiple paths, we can compare them through differences at a certain point in time.

\begin{table}
  \caption{Data summary}
  \label{tab:data}
  \begin{tabular}{ccc}
    \toprule
    Feature               & Source      & Example visual encodings \\   
    \midrule
    time stamp            & time        & 1D space, animation, label \\
    3D position           & motion      & 1D, 2D, or 3D space \\
    3D orientation        & motion      & texture, material, shape \\
    speed of movement     & motion      & color, thickness, texture \\
    speed of rotation     & motion      & color, thickness, texture \\
    pitch / played note   & audio       & color, label \\
    loudness              & audio       & color, thickness, texture \\
    expected note         & sheet music & color, label \\
    expected loudness     & sheet music & color, thickness, texture \\
  \bottomrule
\end{tabular}
\end{table}

\subsection{Visualization}

Having these data features, we can now choose from a range of visual attributes to map them to.
Our current design supports mapping the motion over time to 3D, such that the motion path is ``flying away'' from the user in a direction of choice, which is usually roughly orthogonal to the playing motion (\autoref{fig:time2space}).
The speed of this 3D translation can be adapted. 
When set to zero, the motion is played back exactly as it was recorded (\autoref{fig:one2one}).

To show other data features together with time and space, we change the path's appearance at each time step.
For example, we can vary the color and thickness based on the movement's speed or direction, or the audio's pitch or loudness.
Further ideas for encodings include the path's shape (spiky, smooth, zigzag, ...) or texture and material (stripes, roughness, reflectance, ...), similar to related work~\cite{ware2006whalevis}.

For comparison between multiple recordings or a recording and the current hand motion, we can simply juxtapose or superimpose them (\autoref{fig:comparison}). 
Another interesting option would be to explicitly encode the differences~\cite{gleicher2011visual}, which is non-trivial for time-varying motion data and is left for future work at the moment.
One approach would be a mapping from the Euclidean distance between the expected and actual position to a color.
We could also apply encodings from previous motion guidance research~\cite{yu2020perspective}, such as ``rubber bands'' connecting the actual position to the recommended one.
A mapping to 2D paths in combination with multiples views could show the motion from different perspectives and make comparison easier than the 3D view that suffers from occlusion and perspective distortion, especially when displaying more than two paths.
Such comparative visualizations would allow, for example, to study different playstyles, such as playing a piece in a simplified versus the original way.
Showing these paths side by side should clearly reveal differences.

Besides our visualization, we show a virtual representation of the bow for reference and plan to add a virtual cello.
Sheet music can be displayed virtually as well, for use cases where the head-mounted display is worn during playing.

\section{Conclusion}

We propose a concept for visual immersive analysis of cello players' bow movements and explore different design aspects such as data collection, feature extraction, and visual encodings.
During our design study process~\cite{sedlmair2012design}, we collaborate with a cellist as domain expert.
Our current prototype already supports tracking and visualizing movement and speed in a 3D path's color and thickness.

Future work includes further exploration of the visual design space and an evaluation of effectiveness and usability in different use cases, which we plan to conduct as a user study with a group of cellists.
We further plan to extend our approach to augmented reality, so players and audience are able to see each other and the instruments.

\begin{acks}
Funded by Deutsche Forschungsgemeinschaft (DFG, German Research Foundation) under Germany's Excellence Strategy - EXC 2075 - 390740016, and by Cyber Valley (InstruData project).
\end{acks}

\bibliographystyle{ACM-Reference-Format}
\bibliography{sample-base}





\end{document}